%% file: mythesis.tex
\documentclass[11pt,expanded,copyright]{fsuthesis}
\usepackage{listings}
\usepackage{graphicx}

\title{Speeding Up TestU01\protect\\With the Use of HTCondor}
\author{Joshua Beizer}
\college{College of Arts and Sciences}
\department{Department of Computer Science} 
\manuscripttype{Project}              
\degree{Master of Science}                 
\degreeyear{2016}
\defensedate{August 2$^{nd}$ 2016}


\committeeperson{Michael Mascagni}{Professor Directing Project}
\committeeperson{Xin Yuan}{Committee Member}
\committeeperson{Jeff Bauer}{Committee Member}
\committeeperson{Yu Wang}{Committee Member}

\begin{document}

\frontmatter
\maketitle
\makecommitteepage



\tableofcontents



\begin{abstract}
Testing random number generators is a very important task that, in the resent past, has taken upwards of twelve hours when testing with the current flagship testing suite TestU01. Through this paper we will discuss the possible performance increases to the existing random number generator testing suite TestU01 that are available by offering the executable to an HTCondor pool to execute on. We will see that with a few modifications we are able to decrease the running time of a sample run of Big Crush from about five and a half hours to only five and a half minutes.  We will also see that not only the time taken for the testing to complete is shortened, but also the amount of time the user is unable to use their testing computer is reduced to almost none. Additionally, during this paper we will look at the standard implementation of TestU01 and how it compares with a preexisting Parallel version of TestU01. We will be comparing the performance of the standard version of TestU01 with the parallel version so that we have a performance baseline to test our own modifications against. Finally, this paper will also discuss the use of the distributed computing system HTCondor, and cover some basic instructions related to setting up and using HTCondor. HTCondor is already known to increase the performance of a networked group of computers by allowing super users to utilize additional resources from other lesser user’s idle workstation. We will relate the model recommended for HTCondor to a standard computer lab found at a University and explain why they are the perfect candidates for the system.  
\end{abstract}

\mainmatter

\input chapter1
\input chapter2
\input chapter3
\input chapter4
\input chapter5
\input chapter6
\input chapter7

\input chapter8
\input chapter9
\input chapter10
\input chapter11
\input chapter12

\appendix
\input masterScript
\input makesub
\input superstitch
\input check_q
\input empty
\input restart_slaves
\input reset_slaves





 \bibliographystyle{plain}
 \bibliography{myrefs}

\end{document}

%% file: chapter1.tex
\chapter{Introduction}
	With the increased need for random number generators for use in cryptography, computer simulations, developing Monte Carlo methods, and many more, testing random number generators to ensure 	they meet minimum requirements has never been more important. Although testing is important, it can often be put off due to the lengthy run times required to accurately test the random number generator. Currently the leader in random number generator testing is the TestU01 empirical statistical testing suite which offers multiple tests that range from multiple seconds to multiple hours to run. Through this paper I hope to offer a solution that will both decrease the running time of TestU01 and keep the test as accurate as possible. \cite{mccullough2006review} \cite{suciu2012parallel} \cite{suciu2014parallel}
	
    The idea being tested through this paper is to divide TestU01 into sections and then run each section on an individual computer. To assist in running each test on an individual computer, or at least an individual CPU core, we will be taking a look at High Throughput Condor (HTCondor). In theory, when using HTCondor, the user should be able to decrease the run time of a complete test down to the time of a single sub tests in the TestU01 suite. \cite{mccullough2006review} \cite{suciu2012parallel}
	
    In the following sections we will explore the design and goals related to the project at hand. Next in section 3 we will have an in depth discussion of the current TestU01 testing suite, followed by a discussion of a modified version that has been modified to increase parallel performance of the testing suite. Once we cover the testing suites we will be using, we will discuss HTCondor and how it is used. We will then get to the test set up that will be used for this project, followed by how everything was set up. Next in section 8 we will discuss some of the modifications that were made to both TestU01 and HTCondor to get the project working. Once everything is set up we will discuss the scripts that were written to make running a test as simple as possible. Finally, in sections 10 and 11 respectively, we will discuss the testing done on the system as well as the results of those tests. 

%% file: chapter2.tex
\chapter{Design and Goals}

	The first step in developing a better testing suite for random number generators is to define what we will agree will make a testing suite better. Some of the features that are easy to agree upon for making a testing suite better include decreasing the needed run time, getting accurate values, and being user friendly. 

\section{Design}

The design for this project is fairly straight forward, I will be creating and documenting an HTCondor pool that is able to accept TestU01 as a job. The HTCondor pool will be able to run all of the sections of the TestU01 job simultaneously, reducing the overall time required to run the tests. I will also be  configuring and modifying TestU01 to be able to run on a distributed system to decrease its running time. 

\section{Goals}

	As we have previously stated we have three goals for this project. Goal one is to decrease the run time required to complete a test. The next goal is to achieve the same accuracy of the previous test. Finally, our third goal is to make the test suite more user friendly. Two of our three goals should be easy to measure, since it is possible to measure time with the help of GNU time, and accuracy by comparing the output of multiple runs of a test, however the ease of use is a harder goal to measure. To ensure the goal of ease of use is met, I have managed to simplify running the entire test down to entering one command in a terminal. 

%% file: chapter3.tex
\chapter{TestU01}

TestU01 is currently the most popular testing suite use for determining the validity of a random number generator. When TestU01 was released, it replaced the previous test battery called DIEHARD. Because of how much more demanding TestU01 is for a random number generator to pass, many of the top random number generators that had previously passed DIEHARD were now failing TestU01. At the time TestU01 was released DIEHARD was able to test a random number generator in as little as fifteen seconds, however due to the much more demanding tests of TestU01 it could take upwards of twelve hours to run a single complete test of TestU01’s largest test Big Crush. While TestU01 supports many tests, for this project we will only be focusing on three of them, Small Crush, Crush and Big Crush. \cite{mccullough2006review}

\section{Small, Regular, and Big Crush}

	As I just stated, we will be focusing on three of the tests that are available through TestU01. These three tests include Small Crush, Crush, and Big Crush.	
	Small Crush is the fastest of the three tests to run and is made up of ten smaller tests. With the standard version of TestU01, Small Crush is expected to take anywhere from one to two minutes to finish when running on a Pentium four. 
	Crush is the middle test of the three and one of the more reasonable to run. Not only does Crush offer more tests than small crush at ninety-six tests, it is only ten tests less than Big Crush (however Big Crush runs different, more intense tests).  An average run of Crush is expected to take anywhere from an hour and a half to two hours when running on a Pentium four. 
	Finally, we come to Big Crush, the test that we should be able to see the most improvement out of in this project. Big Crush offers one hundred and six tests to be applied to a random number generator and takes an upward of twelve hours to complete when running on a Pentium four. \cite{mccullough2006review}  \cite{suciu2014parallel} \cite{suciu2012parallel}

\section{Performance}

	The original version of TestU01 was written for and run on single core Pentium four processors. Since then, computer hardware has become more powerful and costs relatively less. Over the years, TestU01 has been maintained and is now a bit better when it comes to overall performance. Because of all the updates, which seem to work similar to the ones in the next section, The current version of TestU01 does not really show any interesting compared to what we will be seeing in the next section. Because of this, I would like to take a look at the reported performance from the paper "A Review of TestU01" by B. D. McCullough. Even though these measurements are outdated, I think it is important to look at them to understand the need for performance increases. I would also like to look at the original reported values so that later we will be able to compare the performance increase of the three systems. \cite{mccullough2006review}\\

    \centerline{\includegraphics[width=10cm]{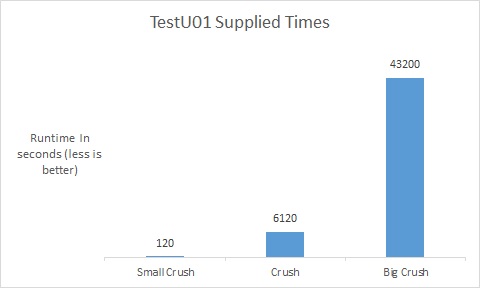}}
    
As you can see, Small Crush takes repetitively no time, requiring only two minutes to run. Next we have standard Crush, which takes about 1.7 hours to complete, making it take long enough for a user to not be productive. Finally we can see Big Crush, which is reported to have taken around 12 hours, which is long enough that a researcher would not be able to use that machine for the rest of the day. As a small disclaimer these are the results reported back when Big Crush only had 45 test vs the 106 it has now. \cite{mccullough2006review}

%% file: chapter4.tex
\chapter{Parallel TestU01}

	A huge breakthrough in my research for this project was discovering that there was an optimized and parallel version of TestU01 that had been created since the release of the standard TestU01 suite. In 2012 the Department of Computer Science at the Technical University of Cluj-Napoca released a project that started out with the standard implementation of TestU01, then first optimized it for modern computers that had access to more cores than they did in the past by increasing its parallel performance. Next they continued and made their parallel version of TestU01 object oriented to increase the efficiency to further reduce the run time of the tests.  \cite{suciu2012parallel} \cite{suciu2014parallel}

\section{Parallel vs. Standard}

Some of the changes the parallel version of TestU01 holds over the standard version as explained by the authors include allowing for multiple input fields to be processed simultaneously, separation of the tests output streams, and finally parallelization using OpenMP. The main feature offered by this implementation that is useful for my project is the parallelization using OpenMP. Essentially, the creators noted that when TestU01 was run it would run through a sequential test of calling the random number generator, then testing, then calling the random number generator, then testing, and so on until the testing was complete. The creators went on to explain that they basically created individual OpenMP sections for every generator call followed by a test. \cite{suciu2012parallel}

Other differences between the two versions of TestU01 are related to how the two get their data for testing. In the original version of TestU01 all of the tests being run were run sequentially. Therefore, they were able to reuse the data from previous tests so that they did not have to get data from the generator every run. The parallel version of TestU01 differs from this because you are able to run each section individually. While the parallel version still supports the option to run sequentially with the same data all the way through, the broken up runs all require their own instances of the random number generator to be run. While this does change the results for an entire run, it does not invalidate them because the same runs are still being run on the generator, it is just being refreshed after each test instead of using the previous generator state. \cite{suciu2012parallel} \cite{suciu2014parallel} \cite{mccullough2006review}

\section{Performance}

	To test the newer parallel system, the creators at Cluj-Napoca were able to run tests using between one and twelve cores to see if there was a point where performance seemed to stop increasing. When the core count was increased from one to three, there was a decrease in time of each single test from about thirty minutes each to about five minutes each. The time to run the tests seemed to stabilize around five minutes per test, with little reduction in run time after the first three cores.  \cite{suciu2012parallel} \cite{suciu2014parallel}
    
	For this project I ran my own tests on this version of TestU01 which can be seen below. I to make the tests fair in comparison to my testing for my own implementation, I ran the tests on one of the machines that will be used in my HTCondor pool. \\
    
   \centerline{\includegraphics[width=10cm]{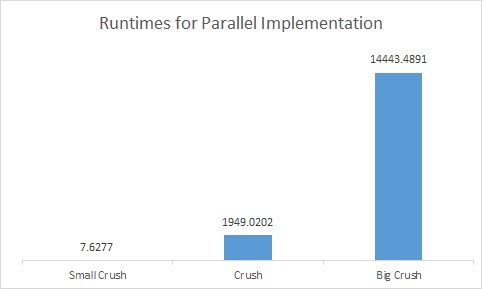}}

	The data above was collected over twenty runs of Small Crush, Crush, and Big Crush. The shortest test, Small Crush, managed to complete in an average of 7.6 seconds compared to the two minutes that was originally quoted before the parallel version of TestU01 was release. The time improvement is very impressive, but Small Crush is not the test that causes issues for a user to test multiple times. As we get to Crush, which was originally quoted as taking around two hours, we see that by just using the parallel version you are able to cut the time to around 32 and a half minutes (1949.0202 seconds). Finally, if we look at the original quote of about 12 hours for Big Crush we can see that the parallel version was able to cut the time down to a little over four hours (14443.4891 seconds).

%% file: chapter5.tex
\chapter{Parallel vs Simultaneous Execution}
	While the main goal of this project is to be able to run the entirety of a TestU01 Big Crush test at once to decrease testing time, it is not exactly the same as running it in parallel. More so, I would like to think of the execution as simultaneous. As stated before, TestU01 is very sequential, with test one finishing, then starting test two, and so on. When running all of the tests at once, you would see no benefit from waiting for each test to complete before starting the next test. While I am utilizing the parallel version of TestU01 that we discussed in the previous section, I am not using it for the parallel functionality, although that is a slight benefit. The main reason I am using the parallel implementation of TestU01 is that it has an already partially working feature that allows you to call individual tests from Crush and Big Crush. Thanks to this, I was able to finish all three of the Crushes and verify the results. \cite{suciu2012parallel}

%% file: chapter6.tex
\chapter{HTCondor}

	The observation that lead to the implementation of HTCondor is that there are two main types of computer users, normal users and power users. Normal users only use a fraction of their computers potential and often times leave their computer idle for long periods of time, or at least they do not fully utilize the hardware they have access to. The other type of user is a power user, or someone who is ultimately limited or bottle necked by their workstation and would benefit from having more resources at their disposal. HTCondor is a way to allow power users access to more resources without spending money, and while increasing the utilization of machinery that is already owned. The core principal of HTCondor is that, if a machine is not currently in use, lets put it to work. A standard install of HTCondor will monitor the usage of all of the computers in the pool and only use the resources of a computer who has been using less than 3\% of its CPU resources and has not seen any mouse or keyboard i/o within the past fifteen minutes. All of the parameters related to when the pool can use a node are able to be changed by the pool manager on a node by node basis to allow users to have more ownership of their computer, while still sharing the idle resources.  While offering resources to power users is considered important, the creators also knew that it would be difficult to get non-power users to allow it on their machines if it would interfere with their own work. In most cases, Condor pools are set up on  large networks, often over two hundred computers. For this project I will only be using nine computers, however since the goal is to allow one user access to a more powerful system, this should be more than adequate. I would also like to point out that when HTCondor was first released each computer would only have one core on average so they only worked as one node. The machines I am using for my experiment all have 4 physical cores with hyper threading, allowing them to be seen as 8 nodes each. Taking that into consideration, I am using about 72 nodes.   \cite{litzkow1988condor} \cite{litzkow1990experience} \cite{wright2001cheap} \cite{gubb2012implementing} \cite{qin2005dynamic} \cite{pruyne1996interfacing} \cite{lodygensky2003xtremweb} \cite{de2003flexible} \cite{karavanic1997integrated} 

\section{Useful Condor Commands}
	During my time working with HTCondor I became acquainted with a few commands that are either useful to a user, or flat out required to get your HTCondor pool to do anything. For a new user, the most important commands are the ones that follow. There are many more commands that a more advanced user may end up using, but those can be found through the online HTCondor documentation. 
    
\textbf{condor\_submit}		submits the specified job to the HTCondor pool. Without this command your HTCondor pool is useless. A user can run this command by entering “condor\_submit submission\_file”. 

\textbf{condor\_q}	will give the user a real-time update of what jobs are in the HTCondor job queue. This command will also show the user if any of the jobs that have been submitted are currently running, being held due to an error, or if they are still idle. This command can be used by entering “condor\_q”. 

\textbf{condor\_release}	allows a user to specify a job number to release from a hold state. If a job is unable to run for any reason, such as incorrect permissions, it will be placed in the hold queue. To release a job from the hold queue the user will want to check the log file to verify why the job was held in the first place, then correct the issue. Once the issue has been resolved, the user can enter “condor\_release Job\#” and the job will go back into the running queue. 

\textbf{condor\_status}		allows the user to view the current status of all of the nodes of the HTCondor pool. The status of the nodes can be unclaimed, claimed, or busy. A node is noted as busy if it is currently being used by another end user, or if the computer has a process running on it that is utilizing more than 3\% of its processor. The node is in the unclaimed state when it is not in use from another end user, and is not currently running any jobs from the HTCondor pool. And finally, a node can be claimed if it is either currently executing, or preparing to execute a job that has been submitted to HTCondor. 

\textbf{condor\_rm} 	allows a user to remove a job from the HTCondor queue before or during its execution. The user is able to run “condor\_rm job\#” to remove any job or part of a job that has been submitted.  

%% file: chapter7.tex
\chapter{Setup}

	The current HTCondor Pool that I have been using for this project is set up in MCH202 and spans nine physical machines. Each machine is configured identically physically with an Intel i7 4770 with a clock of 3.4 GHz (boost clock of 3.9GHz), this CPU is quad core and supports hyper threading allowing to to be used as 8 nodes in the pool. The machine also has  sixteen gigabytes of ram that are accessible by the pool. Software wise, each machine is also identical only differing in host names, with each machine using the name slave\# (\# being the number given to the machine 1\-9, to specify the physical computer).  Later in this paper we will discuss the limitations of this setup, however for our purposes this is adequate. 
	The Master node for this project was created in a virtual machine, that way at the end of the project it is able to be handed over to another user so that they can use this already running pool for their own random number generator testing. The virtual machine is running Linux Mint 17.3, which at the time of creation was the most up to date version of Linux Mint available. Linux Mint was chosen because it seems to be one of the easier distributions for non-Linux users to get equated with, which ties into the goal of making something anyone can use. For the virtual machine it is recommended that the user allow the virtual machine access to at least two CPU cores and at least two gigabytes of ram, that being said it, it should run fine on lower specs. Because the actual execution runs on the remote HTCondor pool, the hardware recommendations are only here to allow the user a good experience with the standard Cinnamon GUI supplied with Linux Mint, and will not affect performance of the HTCondor system. If the end user were to want to create a new master running a different operating system, they would be able to any operating system they would like. They would need to have the host name of master, unless they want to manually change all of the other nodes to link to a new name. To create the new master, the user would just have to install the same version of TestU01 (with modifications) as the rest of the pool, and then install HTCondor and mark their new machine as a submission machine and scheduler. 

\section{Networking}

	Before discussing the steps needed to set up the networking aspect of this project, I would like to bring up the fact that all of the machines intended for use in the HTCondor pool (both execution machines as well as submission machines) must be connected on the same LAN.  If machines are on different LANs it is still possible for a condor pool to be set up, however that is called Flocking, and that is beyond the scope of this project.   \cite{epema1996worldwide} \cite{lodygensky2003xtremweb}
The first step in setting up the HTCondor pool is to configure each machine to be able to communicate with each other through Secure Shell (ssh). Setting this up requires the user to install openssh-server and then generate a public and private key. The command for installing openssh-server is as follows: “sudo apt-get install openssh-server”. Once the software is installed, the user must create their own public and private key. To create keys, the user should navigate to “~/.ssh” and enter “ssh-keygen -t rsa”.  Once all of the machines that are intended to be set up for the HTCondor pool have their own public and private keys, the user needs to disperse the keys. Key dispersal in this case requires n$^{2}$ exchanges so I created a script for the user that will call the command automatically for each transfer on each machine. The manual command the user should use if they need to transfer fewer keys is “ssh-copy-id username@ipaddress”. 
	Now that all of the machines are able to SSH without the need of a password being entered, the user needs to modify the Hosts file to allow them to not have to manually enter the IP address of each machine. The easiest way to accomplish this is to copy the hosts file from the Master node to all of the Slave nodes. The user is able to use the command “sudo scp /etc/hosts username@ip” to send the hosts file to each user manually.  Once the file is on a computer, the user needs to log into the computer and replace its hosts file with the new one that was copied over from the master node. Once the file is copied into place, the user needs to restart networking, which can be done by rebooting. Now that the user is able to ssh from machine to machine using “hostname@hostname” it is time to install HTCondor and build our pool.

\section{Installing HTCondor}

	When installing HTCondor, the user will want to have some information already figured out and close by, such as the names of the computers that will be included in the HTCondor pool, which computers will be for execution, which computers will be allowed to submit a job, and finally which computer will be hosting the HTCondor scheduler. Once the user has all of the information ready, they are able to start the configuration by going to a terminal and entering “sudo apt-get install htcondor”. If the user decides they would like to install on Windows or OSX they are able to download a corresponding executable that will install HTCondor for them. 

\subsection{Standard installation}
	Once the user has started the installation download the user will be greeted with a configuration GUI that walks the user through their install, however without previous research it is easy for a user to get lost during the install. 	The first question asked by the configuration tool is if the user wants to allow the tool to set up HTCondor for them. I strongly recommend the user select the “yes” option here because it drastically simplifies the set up after this point if they do. Next the configuration tool will ask if the user wants to set up a personal HTCondor pool. It is important that the user select “no” at this point, because selecting “yes” will end the set up there and not offer assistance with networking or setting up a scheduler. Next the user will be prompted to select what permissions the current machine will be allotted, whether it is going to be able to submit jobs, execute jobs, and/or be a scheduler (you only need one scheduler). Next the user will be prompted to enter a directory domain label, for this installation just click enter because it is unneeded. This is the point where we will finally see the benefits of setting up our etc/hosts file previously, instead of having to manually type the ip address of the computer that is going to be designated as the scheduler, we are able to just enter the computers name, in our case “master” then click enter. Finally, the user will be prompted to enter the names of all of the computers that will be in the pool and be able to write files to this specific computer. The user will want to enter the name of all of the computers that they will be adding to the pool so that they are able to use check pointing functionality as well as simple functionality like reporting results.  At this point, the installation is almost done and the user will only need to modify two more files. The first file to modify is located at \/etc\/condor\/config.d\/00debconf, where the user will be modifying 3 lines including from the bottom up, ALLOW\_WRITE, CONDOR\_HOST, and finally CONDOR\_ADMIN. The user will want to replace the line that starts with ALLOW\_WRITE with the names of all of the machines in the pool, for example ours says "ALLOW\_WRITE = master, slave1, slave2, slave3, ..., slave9". Next the user will need to change the name of the CONDOR\_HOST to "CONDOR\_HOST = master". Finally, the user will need to change the admin email account encase of issues,the user can just replace the email address with one for the pool to use. The last file the user has to modify is the condor\_config.local located at /etc/condor/ where they need to add the scheduler and other nodes allowed access to it at the bottom. Finally, the user is able to reboot the machine and restart the HTCondor service with “sudo service condor restart”. 
	This set up process will take about five to ten minutes per computer, but is worth it in the long run. Because of the GUI I found it hard to script a solution to configure this step, although I did find it much easier to simply copy the two files over from the master machine once it was configured to all of the slaves and just put them in the right location once the main installation was complete.  To copy the files over the user can use “scp filename compuer@adress:location” or any other method they would prefer.  

\subsection{Parallel Configuration}
	
    While we do not need the parallel configuration for this specific project, it is only a little bit more tweaking of the system and can be beneficial to set up while the user is already in the mindset of configuration. The user will need to go back to /etc/condor/ and add the parallel configuration file (configuration file will be supplied with the master virtual machine). Once the new file is in place the user must specify the new dedicated scheduler in FILENAME with the entry of  ENTRY. When this is all in place the user will be able to restart all of the computers and restart HTCondor once they restart. The nodes should rejoin the pool, and the user will now be able to submit jobs in the parallel universe alongside the vanilla universe. 

\section{Installing TestU01}
	
    Installing TestU01 is a lot easier than installing HTCondor, it only involves moving your TestU01.zip to the /etc/, unzipping it, and then installing the package. A basic installation can be done following the readme that is included in the TestU01.zip file that is available multiple places online. The steps for installing all three versions of TestU01 will be very similar up until the very end, so I will only cover the version that I have modified in this section. The first step the user will want to take is to copy TestU01.zip to their /etc/ directory with “sudo cp TestU01.zip /etc/”. Next the user will want to unzip the folder using “sudo unzip TestU01.zip”. Now that it is unzipped the user will be able to access the directory and all of its files. If the directory is still not accessible the user may need to change the directories permissions with “chmod xxx TestU01” where xxx is the desired level of permissions the user would like to assign. Once inside the new directory, the user will need to run “sudo configure”, “sudo make”, and “sudo make install”. The three commands will install all of the packages required for the user to run TestU01. We are now at the point where the three instillation will differ. For out installation, the user will want to copy the new version of bbattery.c and bbattery.h into their designated directory’s (denoted in the README file supplied with them). At this point the user may wish to reboot all of the systems.  
    
\section{Moving the Pool}
 
 While the HTCondor pool is now located on 9 physical computers in MCH202, it has not always lived there. The current HTCondor pool is actually the third one that was created for this project. Previously the Condor pool was created in MCH202, then recreated as a virtual pool, then moved to physical machines in another lab, and finally found their way back to MCH202.
	As I have stated, the HTCondor pool was originally created in MCH202 towards the beginning of this project. This room seems to be the ideal place for a project of this nature because it has thirty reasonably powerful computers that are idle more often than not. However, as classes started reserving the room, some of the machines were accidentally re-installed over by other people, causing me to have the need to move the system so that this would not happen again.   
The second iteration of the HTCondor pool was created as a single virtual machine that could be cloned and run multiple times. This idea seemed to be a good idea because with this implementation the entire system could be moved anywhere at any time with the creation of a few scripts that would allow the user to easily maintain their networking hosts file as well as their HTCondor networking files. This implantation seemed to be going great until after some research I learned that both HTCondor and TestU01 can sometimes have errors that do not get reported or logged when running on virtual machines. 
After learning about the issues related to running on a virtual machine, and the issues with my previous physical installations being erased accidentally by other users, I was offered the use of forty physical machines located in a more isolated locked computer lab. At first this iteration seemed to be fairly promising and I was able to re-install all of the operating systems and then install and configure HTCondor and TestU01. While installing the operating system I was using a generic Linux 32/64 bit install DVD and did not think to check which version was being installed. After configuring all of the system I attempted to submit a job to my new HTCondor pool and was continuously greeted by a new error I had not previously seen. After researching the error, I realized that the pool was set up on 32-bit Linux and would only accept code compiled for 32bit to run on it. I continued researching and realized that the Pentium 4 processors in the machines were the last generation of Pentium 4 to not support 64 bit operating systems, so I realized once again, I had to move the pool. I do want to note that it would have been possible to compile the code on a 64-bit computer with a flag to allow execution on a 32-bit machine, however that added more complexity to the user’s experience, and I wanted to make sure that there would not be any compatibility issues. 
At this point the semester was well on its way and the state of the computers in MCH202 had been finalized. I was able to create accounts on nine machines that were not being used for a formal class and use them for the remainder of this project.

%% file: chapter8.tex
\chapter{Modifications}
	Most of the work done for this project was either related to figuring out the core installations, moving the system multiple times, and the scripts written to make everything go, all covered in different sections of this paper. There was, however, a chunk of the project that required modifying code for TestU01. 	
	While earlier in this paper we talked about how the parallel implementation of TestU01 had a function to allow the user to call individual tests from any of the three main tests (Small Crush, Crush, or Big Crush), I did not mention that the functionality was never directly used anywhere. To be able to run tests individually, I had go though and figure out how the single calls worked. The preset individual tests also only worked on Crush and Big Crush, so I had to manually get it to support small Crush also. As it was when I first touched the parallel version of TestU01, when Small Crush was run, tests 3 4 and 7 were never run. After fixing some dependencies I was able to get them working again. 
	The main fix for getting Small Crush working as individual tests was to change bbattery.h and bbattery.cpp to accept an extra parameter for the “smallcrush” function. The extra parameter was used to pass in which of the ten tests the user wants to run for small Crush. The other two tests were set up to handle selection of tests slightly differently, however the implementation ended up being very similar, but more complicated. Other modifications in this project manifested in the form of configuration modifications discussed in other sections of this paper. 

%% file: chapter9.tex
\chapter{Scripts}

	While the configuration of this project is fairly nontrivial, I wanted its use to be simple enough that it could immediately benefit anyone testing a random number generator with little to no instructions. To achieve this goal, I created scripts that only require a user to be able to compile their own code and then run a single script.   To achieve this goal, I created seven scripts that are able to run different tasks related to using TestU01 and HTCondor. Below I will explain in detail what each script does and when it would be run.
    
\section{HTCondor Related Scripts}
\textbf{check\_q} 	is a simple bash script written to constantly monitor the job queue for HTCondor. While this is the shortest script, it is still very useful for monitoring the status of longer jobs. When the script is run it automatically calls “condor\_q” every two seconds. This script was written to help the user monitor when the HTCondor pool accepts the submitted job, starts running the submitted job, and the state of the running job.
Run by ./check\_q

\textbf{restart\_slaves} 	is an expect script that warm boots (restarts) all of the computers in the condor pool. This script can be used after a configuration change is implemented on some or all of the execution machines, such as modifying the networking hosts file. 
Run by ./restart\_slaves 'password'

\textbf{reset\_slaves} 		is an expect script that restarts the HTCondor service on all of the nodes in the condor pool. This script is used after nodes in the pool are restarted, disabled, or otherwise not connecting to the HTCondor scheduler. 
Run by ./reset\_slaves 'password'

\textbf{makesub} 	is an automated bash script that creates the HTCondor submission file for the user. This script requires the user to have little to no experience with HTCondor, but is easily modified to allow for advanced configurations. In its current state makesub will only create a submission file that uses the vanilla universe, but can be changed my manually editing the file to support other universes. Makesub takes two parameters, the executable file and the name of the test being run. The output file will differ depending on the test suite being run. This script is able to be run individually if the user wishes to create or save their submission file for other uses. 
Run by ./makesub executable test\_name

\section{TestU01 Related Scripts}
\textbf{empty} 		is a bash script that was created to determine when HTCondor is finished running a job and to check to make sure all output files were generated. If the script completes and not all of the files are generated, it will give an exit status of 1 and echo how many of the files were correctly generated so far. This script works by checking the directory for files named output.\# with \# being the number of the process. When empty finds a file it will check its size to see if it is larger than 0. If the file exists and is larger than 0 it will increment the program counter in empty. The overall script will know all of the files are created and valid by comparing the amount it found to be correct with the number of files that should be created for the specific test (ex. Small Crush only needs 11 files while Big Crush needs 107 files).
Run by ./empty test\_name

\textbf{Superstitch}		is a bash script that takes all of the output files generated by the test run and joins them together into one results file (results.txt). This results file is then copied to the directory of the users’ choice if the directory exists. If the directory does not exist, it is created and then the file is copied into it. This script also moves all of the output files into the same directory to clean up the users current working directory while saving previous test results in their original form. This script can be run independent of the Master script if the user would like to check old files from previous runs that they need a new results file for. 
Run by ./superstitch test\_name directory

\textbf{master} 	is the main script that automates the test for a user who may or may not know anything about HTCondor and very little about TestU01. This one script will take the user from start to finish with their test only requiring them to supply an executable while asking them where they want to save the results and which test they would like to run. Below is an example of the output generated by this script as it runs. 
 
When “master” is run, line 1 will immediately be generated, followed by the script calling “makesub”. Once “makesub” has finished a file called “sub” will be generated, followed by “master” displaying that the job is being submitted to HTCondor. Next “master” will submit the job to the HTCondor pool with “condor\_submit sub” and prints the condor cluster number. The cluster number can be thought of as a priority number, where lower numbers are executed before higher numbers (as possible).  Once the job has been submitted, the “master” script goes into a loop that is terminated only when all of the files have been properly generated by HTCondor. So that the loop does not use more system resources than it really needs, it only checks the status of the generated files every twelve seconds. A file is deemed to be correct once the file size is greater than 0, this is done because all of the files are actually generated empty once HTCondor accepts the job, but will only write to a specific file when its corresponding test has completed.  Once all of the tests are deemed complete, “master” will call “superstitch” which copies the content of all of the generated files into one easy to read “results.txt” file and gives the user feedback on its progress by printing the file number that it is currently working on copying.  Finally, “master” moves all of the files that were generated to the directory that was specified by the user at the beginning of the run, copies the results file to the directory specified by the user, and cleans up the current working directory of any files that were created, leaving the user with a directory in the same state as when they started. 
Run by ./master executable test dir 

%% file: chapter10.tex
\chapter{Testing}
 
	When it came to bench marking and validating the HTCondor pool running TestU01 I had a few measurements I wanted to address. The first important metric of the system that should be tested is the speed increase, which can be measured by running the same tests on all three versions of TestU01 on the same hardware while using the GNU Time command. To ensure I am not just getting a particularly good or bad test run when testing, I have decided to run each test twenty times on each system.  The second important metric to test is accuracy of the test. To do this I have used GNU diff to compare the output of the three versions of TestU01 to see if they all agree with one another. Obviously, when looking for differences, we are able to ignore time differences since they are not related to accuracy. Finally, I took a look into the usability of the machine that submits the TestU01 test. For doing this I monitored CPU usage on the client machine while running the test. Obviously there are other factors related to a good user experience, but if the user is unable to use a chunk of their CPU all other tasks will also be negatively affected. 

%% file: chapter11.tex
\chapter{Results} 
The results of my tests are fairly not surprising, when more work can be done at once, it gets done faster. During my testing of the HTCondor pool running TestU01 I limited the number of cores down to forty cores so that I could ensure the machines being used were not touched while being tested. As I have stated previously, this HTCondor pool is set up in an actively used computer lab, and some of the computers are used by students daily. After running tests on different numbers of cores, I found that it is fairly simple to figure out an approximation for how long a test will take to run. TestU01 has 106 tests to run, if you have 40 cores for it to run on, it will have to run the test in 3 batches. If each batch takes about 4 minutes, you can expect the test to take about 12 minutes total. If you look at the following stats, you can see that the average run for Big Crush was approximately 11 minutes. Similarly, I saw that if I raised the core count to 70 it would require only two batches of tests, which would cut the run time by about four minutes.  Finally, If I ran all 90 cores That I had access too, it would still require 2 batches, so it would not decrease in time. I decided to keep my results with 40 cores to show more of a worst case scenario. \\

\centerline{\includegraphics[width=10cm]{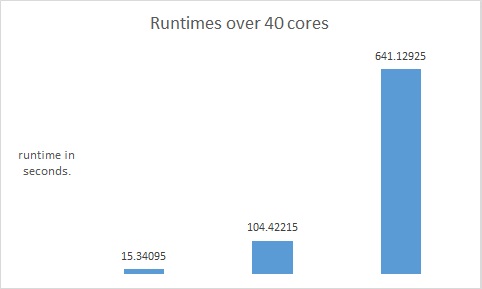}}

As you can see, the overall times look to be in similar proportions to the tests done to the parallel version of TestU01, however most of my numbers are significantly smaller. Below are the results for all of the tests narrowed down to the specific test, and compared to the parallel implementation. \\

\textbf{Small Crush}\\

\centerline{\includegraphics[width=10cm]{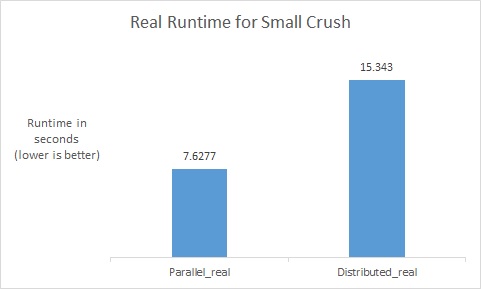}}

As you can see by the previous chart, the average time taken to run Small Crush actually increased during my tests due to the overhead related to submitting and negotiating a job with HTCondor. When running Small Crush it may be more useful for the tester to use a non modified version of TestU01. That being said, the difference between almost 8 seconds and 16 seconds, while realistically being double the time, is not really long enough to make much of a difference in a users day. 

\centerline{\includegraphics[width=10cm]{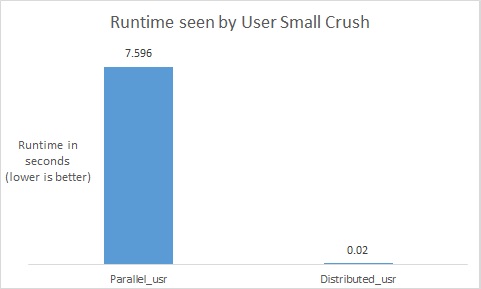}}

Above is a chart that shows the average amount of time the users personal computer is in use for the testing of their random number generator. On all of the tests offered by TestU01, the CPU usage of the test will be 100\% of the cores allowed to run the test. Even though sending small crush to the HTCondor pool has more overhead and ends up taking longer, it will only pin the users cpu at 100\% for an average of .02 seconds, compared to the 8 seconds taken by the standard version. That means that if the user has any other tasks running on their machine, they will not be effected nearly as harshly as the standard and parallel versions. \\

\textbf{Crush}\\

\centerline{\includegraphics[width=10cm]{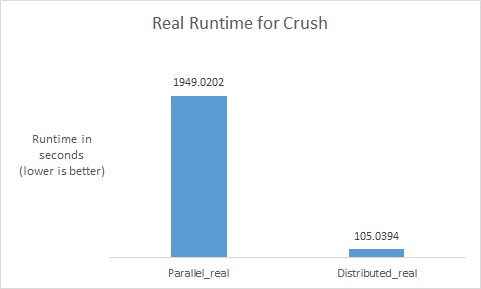}}

As you can see by the graph above, standard Crush is really where this system starts to shine. While the parallel version of TestU01 takes an average of 1949.0202 seconds (about 32 and a half minutes) to run, by submitting it to HTCondor we are able to save 1843.9808 seconds ( about thirty-one minutes) leaving us with only 105.0394 seconds ( 1 minute and 45 seconds). I think it is very agreeable that if a user is able to run the same test in about two minutes (the starting length of Small Crush), they would be able to get much more work done in a single day. 

\centerline{\includegraphics[width=10cm]{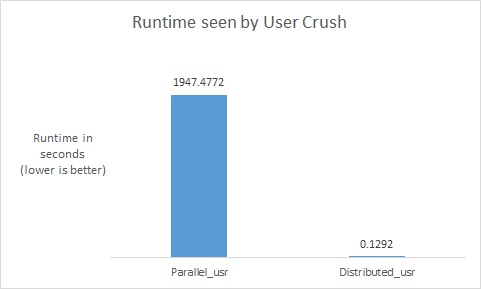}}

Just like with Small Crush, we see the similar results when it comes to the time difference between how much time the user will be effected by running the tests. As with Small Crush, if the user were to use the parallel version on TestU01 only, they would have their CPU pinned to 100\% utilization for the full 32 minutes it would take to run their tests. However, if the user were to submit their job to the HTCondor pool, they would see only .1292 seconds of CPU usage, leaving their other tasks free to run in the background. \\

\textbf{Big Crush}\\

\centerline{\includegraphics[width=10cm]{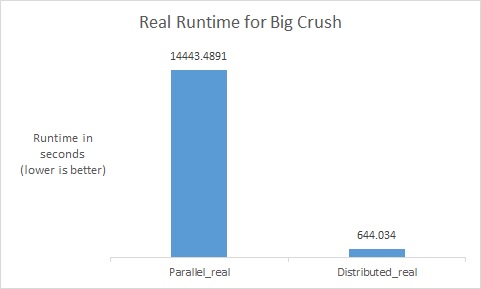}}

Finally we can take a look at Big Crush, which has the largest room for improvement time wise. While the increase from the estimated 12 hours down to about four hours by the parallel implementation is pretty impressive, we are able to get the time down even further to about 10 minutes by using the HTCondor pool. As you can see we have an average run time of 644.034 seconds for a full run of Big Crush (which works out to about 10 minutes and 45 seconds). As I stated previously, the decrease from 12 hours to 4 is great, but now the user is able to run the entire test in a matter of minutes, saving them more time than any other test. 

\centerline{\includegraphics[width=10cm]{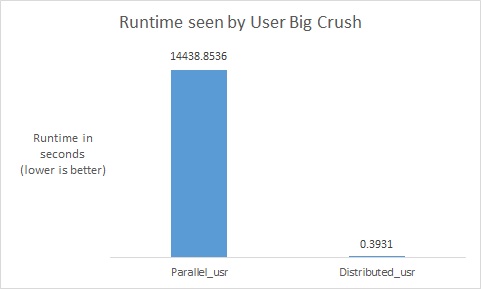}}

The best part about this performance change is that not only does the test take significantly less time to run, but the user is able to use their computer while they wait for the results. With the test running locally through the parallel version of TestU01, the user would see their CPU stuck at 100\% utilization for the full four hours. With this version however, we are able to get the users noticeable time down to .3931 seconds, leaving the user able to do whatever they would like while the test runs. \\

\textbf{Accuracy}\\
Finally, one of the most important factors to this entire testing suite is the accuracy of the tests. As I explained in the previous section, to test the accuracy of the test runs, I compared all twenty runs of each test against themselves. The results of the diff testing was always the same between runs of the same type, the only thing that differed was the time each test took to run. Once I was happy with all of the values that I found based off of the twenty runs of each, I was able to diff one run of each type (since all of the runs for each test were the same as each other). The results of all of the HTCondor submitted tests were identical to the tests given by the parallel version of TestU01 run individually. That being said, the results of my tests were different from the results from the standard sequential run of the parallel version of TestU01. The differences between the runs boils down to the fact that, as we talked about earlier, when TestU01 is run sequentially it uses previous generator values in the following test. In the individual tests offered by the Parallel version, instead of using the value from the previous test, it calls the random number generator again to get new values to work with. While some of the result values are different between the sequential tests and the individual tests, they are still testing in the same method, however they are testing different values, which should come up with different results. According to papers related to the parallel version of TestU01, the values are still legitimate values. \cite{suciu2012parallel} \cite{suciu2014parallel}

%% file: chapter12.tex
\chapter{Conclusion}
	Testing random number generators has been a lengthy but important part of random number generator development for a long time, however we finally have a solution that not only has the accuracy of the flagship testing suite TestU01, but it also takes significantly less time to run. Through this paper, we have discussed three different implementations of TestU01, along with how to set up and use an HTCondor pool. By utilizing multiple computers at once we were able to reduce the time it takes for TestU01 Big Crush from four hours to ten minutes which allows users to use their time for other things, while also keeping the accuracy of the test compared to the results of previous implementations of TestU01.  
	Some future research that could be done to further the validity of this project would be to fully prove that the resulting values are one hundred percent accurate and valid. At the moment, the validity of the results hinges on the fact that they match the published results of another research group. The main issue with proving the validity of the tests is that there are over 100 tests to go through to figure out if the values related to each test, which ends up being out of the scope of this project. 
    While this work is by no means the final say in increasing the performance of TestU01, it is a big jump in accessibility of the suite. Because of the decreased run time users are able to test their random number generators much faster, and then modify them and retest them same day, which is not as possible with the test taking multiple hours. This method of testing is also a lot more user friendly than previous versions due to the included scripts and other files that automate the testing suite for the user.

%% file: masterScript.tex
\chapter{Master}
\begin{lstlisting}
#!/bin/bash

####################################
# Created by Joshua Beizer         #
# this script is run to automate   #
# the entire testing process.      #
# When the script is run it starts #
# by running makesub to create the #
# required submission file. Next   #
# submits the file to the HTCondor #
# pool to start the execution of   #
# the test. At this point it goes  #
# into a loop of waiting for the   #
# tests to be done. It checks for  #
# this by running empty. in the    #
# event of a hold, the script fixes#
# the issue and then releases the  #
# hold. when everything is complete#
# the script will join everything  #
# using superstitch. Finally, the  #
# script cleans up any tmp files   #
#that were generated during testing#
####################################
# Notes:                           #
# use: ./master executable test dir#
#ex : ./master test.out crush test8#
####################################

InputFile=$1
Battery=$2
Folder=$3
NotGenerated=0

echo "making HTCondor submit file"
  ./makesub $InputFile $Battery > runTest
  sleep 2

echo ""
echo "submitting to HTCondor"
  condor_submit runTest > tmp2
  echo -n "cl=" > tmp3
  grep -oP "cluster\s+\K\w+" tmp2 >> tmp3

  #pull value from tmp file 3 (just cluster number)
  source tmp3
  ClusterNumb=$cl
  echo "Condor Cluster Number : $ClusterNumb"
  echo ""

  echo "checking for output files"
  echo "This may take some time, files are written after"
  echo "their corresponding tests have finished running"
while [ $NotGenerated -eq 0 ]
do
  ./empty $Battery 
  if [ $? -eq 0 ]
  then
    let NotGenerated=1
    echo "files generated"
  else
  condor_q > tmp
  echo -n "held=" > tmp4
  grep -oP "running,\s+\K\w+" tmp >> tmp4
  source tmp4
  hVal=$held

  if [ $hVal -ne 0 ]
  then 
    chmod 777 output.*
echo ""
    condor_release $ClusterNumb 
    echo "held tests released"
  fi
  sleep 3
  echo -n " ."
  sleep 3
  echo -n " ."
  sleep 3
  echo " ."
  sleep 3

  fi
done

echo ""
echo "Joining all output files"
  ./superstitch $Battery $Folder

echo "files joined, results.txt generated"
echo "cleaning up directory"
  rm runTest
  rm tmp
  rm tmp2
  rm tmp3
  rm tmp4

echo ""
echo "*******************************************************"
echo "Testing complete. Results are in results.txt which is "
echo "located in your current directory as well as in $Folder"
echo "*******************************************************"
\end{lstlisting}

%% file: makesub.tex
\chapter{makesub}
\begin{lstlisting}
#!/bin/bash
####################################
# Created by Joshua Beizer         #
# this script is run to create     #
# the submission file needed to    #
# submit a job to a condor pool    #
####################################
# Notes:                           #
#use:./makesub executable testname #
#ex: ./makesub testfile crush      #
####################################

counter=0
InputFile=$1
Battery=$2
NumbTests=0

if [ "$Battery"  == "SmallCrush" ] || [ "$Battery" == "smallcrush" ]
then
  NumbTests=10
  SelectedTest=0
fi

if [ "$Battery" == "Crush" ] || [ "$Battery" == "crush" ]
then
  NumbTests=96
  SelectedTest=1
fi

if [ "$Battery" == "BigCrush" ] || [ "$Battery" == "bigcrush" ]
then
  NumbTests=106
  SelectedTest=2
fi 

echo "Universe = vanilla"
echo "Executable = $InputFile"
echo "Log = log"
echo "Output = output.\$(Process)"
echo ""

while [ $counter -le $NumbTests ];
do 
  echo "Arguments = $counter $SelectedTest"
  echo "Queue"
  echo ""
  let counter=counter+1
done
\end{lstlisting}

%% file: superstitch.tex
\chapter{superstitch}
\begin{lstlisting}
#!/bin/bash

####################################
# Created by Joshua Beizer         #
# this script is run to create     #
# a single results file for the    #
# user to be able to look at.      #
# when the a TestU01 job is sent   #
# to HTCondor it will result in    #
# 11 to 107 flies depending on the #
# test being run. This script      #
# cleans that up                   #
####################################
# Notes:                           #
# use ./superstitch testname dir   #
# ex ./superstitch crush results2  #
####################################

counter=0
Battery=$1
NumbTests=0
stitchNumb=0

if test -d $2
 then
   echo "directory exists"
 else
   mkdir $2
fi

#set number of files to work on
if [ "$Battery"  == "SmallCrush" ] || [ "$Battery" == "smallcrush" ]
 then
   NumbTests=10
   stitchNumb=1
   touch stats.txt
fi

if [ "$Battery" == "Crush" ] || [ "$Battery" == "crush" ]
 then
   NumbTests=96
   stitchNumb=2
fi

if [ "$Battery" == "BigCrush" ] || [ "$Battery" == "bigcrush" ]
 then
   NumbTests=106
   stitchNumb=2
fi

#remove previous files
rm results.txt
rm stats.txt

#get header 
head -6 output.0 > results.txt

#output.0 is null in big and regular crush, so ignore it
if [ $stitchNumb -eq 2 ]
 then
   let counter=counter+1
fi

#only output.0 has a header in smallcrush so only rm that one
if [ $stitchNumb -eq 1 ]
 then 
   tail -n +5 output.0 > output.0.tmp
   cp output.0.tmp output.0
   rm output.0.tmp
fi

while [ $counter -le $NumbTests ];
do

  if [ $((counter % 11)) -eq 0 ] && [ $counter -ne 0 ]
  then
    echo ""
  fi

echo -n  "$counter "

  if [ $stitchNumb -eq 2 ]
   then
    #drop header from output files 
    tail -n +7 output.$counter >  output.$counter.tmp

    #find where results start  
    lineno=$(grep -n "Summary" output.$counter.tmp | cut -f1 -d:)
   # echo $lineno

    #+4 removes blank space
    let lineno=lineno+4
    echo "$counter" >>stats.txt 
    tail -n +$lineno output.$counter >> stats.txt
    echo "" >> stats.txt

    let lineno=lineno-8 
    # -8 removes some blank lines
    #copy everything but results to file
    head -$lineno output.$counter.tmp >>results.txt

    #rm output.$counter
    rm output.$counter.tmp
    mv output.$counter $2/
   else
    cat output.$counter >> results.txt
    mv output.$counter $2/
  fi

  let counter=counter+1
done
echo ""

#clean up for next round
cp results.txt $2/
cp stats.txt $2/
mv log $2/

\end{lstlisting}

%% file: check_q.tex
\chapter{check\_q}
\begin{lstlisting}

#!/bin/bash
####################################
# Created by Joshua Beizer         #
####################################
# This was made to check the queue #
# for longer running jobs to see   #
# when they switch from idle to    #
# running.                         #
####################################

while [ 1 ]
do
condor_q
sleep 2
done
\end{lstlisting}

%% file: empty.tex
\chapter{empty}
\begin{lstlisting}
#!/bin/bash

####################################
# Created by Joshua Beizer         #
# this script is run to decide     #
# when HTCondor is finished with   #
# a job. It decides it is done by  #
# checking all of the output files #
# and ensuring they are compelete  #
####################################
# Notes:                           #
#use: ./empty testname             #
#ex: ./empty crush                 #
####################################

Battery=$1
var=0
counter=0
NumberDone=0

if [ "$Battery"  == "SmallCrush" ] || [ "$Battery" == "smallcrush" ]
then
  NumbTests=11
fi

if [ "$Battery" == "Crush" ] || [ "$Battery" == "crush" ]
then
  NumbTests=97
fi

if [ "$Battery" == "BigCrush" ] || [ "$Battery" == "bigcrush" ]
then
  NumbTests=107
fi

while [ $var  == 0 ]; do
  if [ -f output.$counter ];
  then
    for filename in output.$counter;
    do
       if [ -s $filename ]
       then
         let NumberDone=NumberDone+1
	fi
       let counter=counter+1
    done
   # FILERANGE = $FILERANGE"?"
  else
    let var=var+1
  fi
done

if [ $NumberDone -eq $NumbTests ]
then 
    exit 0
else
  finished=$NumberDone/$NumbTests
  echo -n "$finished files generated"
    exit 1
fi

done

\end{lstlisting}

%% file: restart_slaves.tex
\chapter{restart\_slaves}
\begin{lstlisting}
#!/usr/bin/expect -f

####################################
# Created by Joshua Beizer         #
# this script is run to            #
# reboot all of the                #
# execution machines               #
# (ie. slave1 - 9)                 #
####################################
# Notes:                           #  
# some chars will not be accepted  #
# normally  from std in (ex. !)    #
# fix by putting password in       #
# single quotes                    #
# ex ./restart_slaves 'password!!' #
####################################

set timeout 120
set password [lindex $argv 0]

for {set count 1} {$count < 10} {incr count 1} {
spawn ssh slave$count@slave$count
expect "Wel*"

send -- "sudo shutdown -r now\r"
expect "*?assword*"
send -- "$password\r"
expect eof
}
\end{lstlisting}

%% file: reset_slaves.tex
\chapter{reset\_slaves}
\begin{lstlisting}
#!/usr/bin/expect -f

####################################
# Created by Joshua Beizer         #
# this script is run to reset      #
# the condor service on all of the #
# execution machines               #
# (ie. slave1 -9)                  #
####################################
# Notes:                           #  
# some chars will not be accepted  #
# normallly from std in (ex. !)    #
# fix by putting password in       #
# single quotes                    #
# ex ./reset_slaves 'password!!'   #
###################################

set timeout 120
set password [lindex $argv 0]

for {set count 1} {$count < 10} {incr count 1} {
spawn ssh slave$count@slave$count
expect "Wel*"

send -- "sudo service condor restart\r"
expect "*?assword*"
send -- "$password\r"
sleep 2
send -- "exit\r"
expect eof
}
\end{lstlisting}

%% file: mythesis.bbl
\begin{thebibliography}{10}

\bibitem{de2003flexible}
J~Chassin de~Kergommeaux and Benhur de~Oliveira~Stein.
\newblock Flexible performance visualization of parallel and distributed
  applications.
\newblock {\em Future Generation Computer Systems}, 19(5):735--747, 2003.

\bibitem{epema1996worldwide}
Dick~HJ Epema, Miron Livny, Ren{\'e} van Dantzig, Xander Evers, and Jim Pruyne.
\newblock A worldwide flock of condors: Load sharing among workstation
  clusters.
\newblock {\em Future Generation Computer Systems}, 12(1):53--65, 1996.

\bibitem{gubb2012implementing}
David Gubb, Violeta Holmes, Ibad Kureshi, Shuo Liang, and Yvonne James.
\newblock Implementing a condor pool using a green-it policy.
\newblock 2012.

\bibitem{karavanic1997integrated}
Karen~L Karavanic, Jussi Myllymaki, Miron Livny, and Barton~P Miller.
\newblock Integrated visualization of parallel program performance data.
\newblock {\em Parallel Computing}, 23(1):181--198, 1997.

\bibitem{litzkow1988condor}
Michael~J Litzkow, Miron Livny, and Matt~W Mutka.
\newblock Condor-a hunter of idle workstations.
\newblock In {\em Distributed Computing Systems, 1988., 8th International
  Conference on}, pages 104--111. IEEE, 1988.

\bibitem{litzkow1990experience}
Mike Litzkow and Miron Livny.
\newblock Experience with the condor distributed batch system.
\newblock In {\em Experimental Distributed Systems, 1990. Proceedings., IEEE
  Workshop on}, pages 97--101. IEEE, 1990.

\bibitem{lodygensky2003xtremweb}
Oleg Lodygensky, Gilles Fedak, Franck Cappello, Vincent Neri, Miron Livny, and
  Douglas Thain.
\newblock Xtremweb \& condor: sharing resources between internet connected
  condor pool.
\newblock In {\em Cluster Computing and the Grid, 2003. Proceedings. CCGrid
  2003. 3rd IEEE/ACM International Symposium on}, pages 382--389. IEEE, 2003.

\bibitem{mccullough2006review}
BD~McCullough.
\newblock A review of testu01.
\newblock {\em Journal of Applied Econometrics}, 21(5):677--682, 2006.

\bibitem{pruyne1996interfacing}
Jim Pruyne and Miron Livny.
\newblock Interfacing condor and pvm to harness the cycles of workstation
  clusters.
\newblock {\em Future Generation Computer Systems}, 12(1):67--85, 1996.

\bibitem{qin2005dynamic}
Xiao Qin and Hong Jiang.
\newblock A dynamic and reliability-driven scheduling algorithm for parallel
  real-time jobs executing on heterogeneous clusters.
\newblock {\em Journal of Parallel and Distributed Computing}, 65(8):885--900,
  2005.

\bibitem{suciu2012parallel}
Alin Suciu, Radu~Alexandru Toma, and Kinga Marton.
\newblock Parallel implementation of the testu01 statistical test suite.
\newblock In {\em Intelligent Computer Communication and Processing (ICCP),
  2012 IEEE International Conference on}, pages 317--322. IEEE, 2012.

\bibitem{suciu2014parallel}
Alin Suciu, Radu~Alexandru Toma, and Kinga M{\'a}rton.
\newblock Parallel object-oriented implementation of the testu01 statistical
  test suites.
\newblock In {\em Intelligent Computer Communication and Processing (ICCP),
  2014 IEEE International Conference on}, pages 311--315. IEEE, 2014.

\bibitem{wright2001cheap}
Derek Wright.
\newblock Cheap cycles from the desktop to the dedicated cluster: combining
  opportunistic and dedicated scheduling with condor.
\newblock In {\em Conference on Linux clusters: the HPC revolution}, 2001.

\end{thebibliography}
